\pdfoutput=1
\documentclass[epj]{webofc}
\usepackage[utf8]{inputenc}
\usepackage[varg]{txfonts}   
\usepackage{booktabs}
\usepackage{xcolor}
\definecolor{darkred}{rgb}{0.4,0.0,0.0}
\definecolor{darkgreen}{rgb}{0.0,0.4,0.0}
\definecolor{darkblue}{rgb}{0.0,0.0,0.4}
\usepackage[bookmarks,linktocpage,colorlinks,
    linkcolor = darkred,
    urlcolor  = darkblue,
    citecolor = darkgreen]{hyperref}
\usepackage{microtype}
\usepackage{mathtools}
\usepackage{fixmath}
\usepackage{scalerel}
\usepackage{dsfont}
\usepackage{xfrac}
\usepackage{physics}
\usepackage{mleftright}
\mleftright 
\usepackage{tikz}
\usetikzlibrary{decorations.pathreplacing}
\usetikzlibrary{patterns}
\usepackage[outline]{contour}
\contourlength{1.5pt}
\usepackage{siunitx}
\wocname{EPJ Web of Conferences}
\woctitle{Lattice2017}
\newsavebox{\foobox}
\newcommand{\slantbox}[2][\slantvalue]{\mbox{%
        \sbox{\foobox}{#2}%
        \hskip\wd\foobox
        \pdfsave
        \pdfsetmatrix{1 0 #1 1}%
        \llap{\usebox{\foobox}}%
        \pdfrestore
}}
\newcommand\unslant[2][-.25]{%
  \mkern1mu%
  \ThisStyle{\slantbox[#1]{$\SavedStyle#2$}}%
  \mkern-1mu%
}
\NewDocumentCommand\I{}{\mathrm{i}}                           
\NewDocumentCommand\e{ m }{\mathrm{e}^{#1}}                   
\NewDocumentCommand\cpi{}{\unslant{\pi}}                      
\NewDocumentCommand\complex{}{\mathds{C}}                     
\NewDocumentCommand\integers{}{\mathds{Z}}                    
\NewDocumentCommand\Id{}{\mathds{1}}                          
\NewDocumentCommand\SN{}{\(\mathsurround=0pt\text{S}/\text{N}\)}
\DeclareSIUnit\fm{\femto\metre}
\DeclareSIUnit\MDU{MDU}
\NewDocumentCommand\D{ l g }{                                   
	\IfNoValueTF{#2}
		{\mathcal{D}}
		{\mathinner{\mathcal{D}\argopen#1[#2\argclose#1]}}
	}
\begin{document}
%
\selectlanguage{english}
\title{%
Local multiboson factorization of the quark determinant
}
\author{%
\firstname{Marco}    \lastname{Cè}\inst{1,2,3}\fnsep\thanks{Speaker, \email{marco.ce@uni-mainz.de}, HIM-2017-06, CERN-TH-2017-221, DESY-17-167} \and
\firstname{Leonardo} \lastname{Giusti}\inst{4,5,6} \and
\firstname{Stefan}   \lastname{Schaefer}\inst{7}
}
\institute{%
Helmholtz-Institut Mainz, Johannes Gutenberg-Universität, Staudingerweg 18, 55128 Mainz, Germany
\and
Scuola Normale Superiore, Piazza dei Cavalieri 7, 56126 Pisa, Italy
\and
INFN, sezione di Pisa, Largo B. Pontecorvo 3, 56127 Pisa, Italy
\and
Theoretical Physics Department, CERN, Geneva, Switzerland
\and
Dipartimento di Fisica, Università di Milano-Bicocca, Piazza della Scienza 3, 20126 Milano, Italy
\and
INFN, sezione di Milano Bicocca, Piazza della Scienza 3, 20126 Milano, Italy
\and
John von Neumann Institute for Computing (NIC), DESY, Platanenallee 6, 15738 Zeuthen, Germany
}
\abstract{%
  We discuss the recently proposed multiboson domain-decomposed factorization of the gauge-field dependence of the fermion determinant in lattice QCD. In particular, we focus on the case of a lattice divided in an arbitrary number of thick time slices. As a consequence, multiple space-time regions can be updated independently. This allows to address the exponential degradation of the signal-to-noise ration of correlation functions with multilevel Monte Carlo sampling. We show numerical evidence of the effectiveness of a two-level integration for pseudoscalar propagators with momentum and for vector propagators, in a two active regions setup. These results are relevant to lattice computation of the hadronic contributions to the anomalous magnetic moment of the muon and to heavy meson decay form factors.
}
\maketitle
\section{Introduction}\label{sec:intro}

It is well known that the standard Monte Carlo (MC) evaluation of a generic multi-point function of QCD has an exponential signal-to-noise ratio (\SN) problem. Namely, the signal decays with the distance between points with a faster exponential rate than its statistical error, or noise~\cite{Parisi:1983ae,Lepage:1989hd}. This currently limits the obtainable statistical accuracy at long distances. A way to tackle the signal-to-noise problem that involves a modification of the MC integration procedure, the so-called \emph{multilevel} algorithm~\cite{Parisi:1983hm,Luscher:2001up}, is briefly introduced in Section~\ref{sec:multilevel}.

Algorithms of this kind have been applied to lattice Yang--Mills theory~\cite{Parisi:1983hm,Luscher:2001up,Meyer:2002cd,DellaMorte:2007zz,DellaMorte:2008jd,DellaMorte:2010yp}, thanks to the fact that the action of a purely bosonic theory is \emph{manifestly local}. However, in numerical lattice QCD the Grassmann quark fields are first integrated out analytically. This leads to a \emph{non-local} fermion determinant and \emph{non-local} Wick contractions. Both non-localities prevent a straightforward application of multilevel sampling.

In the first part of these proceedings (Sec.~\ref{sec:mbddhmc}), we present the factorization introduced in Refs~\cite{Ce:2016ajy,Giusti:2017ksp}, focusing on the case of \emph{independent updates} of an \emph{arbitrary number} of thick-time-slice regions. To this purpose, the fermion determinant is factorized in contributions that are localized to a spacetime region of a domain decomposed (DD) lattice~\cite{Luscher:2005rx}, while interactions between different domains are mediated by multiboson (MB) fields~\cite{Luscher:1993xx,Borici:1995np}. In the second part of these proceedings (Sec.~\ref{sec:numeric}) we show new numerical evidence that multilevel sampling can solve the \SN\ problem for the connected pseudoscalar correlator with momentum and for the vector correlator.

\section{The signal-to-noise ratio problem}\label{sec:snrp}
As a prototypical manifestation of the \SN\ problem, consider the mesonic correlator on a fixed background gauge field configuration
\begin{equation}\label{eq:precorr}
  W_\Gamma(x_0,y_0,\vb{p}) = \frac{1}{L^3} \sum_{\vb{x},\vb{y}} \e{\I \vb{p}\cdot(\vb{x}-\vb{y})} \ev{ \bar{d}(y)\Gamma u(y) \; \bar{u}(x)\Gamma d(x) }_f
  = -\frac{1}{L^3} \sum_{\vb{x},\vb{y}} \e{\I \vb{p}\cdot(\vb{x}-\vb{y})} \Tr{ \Gamma \gamma_5 S(x,y)^\dagger \gamma_5 \Gamma S(x,y) } ,
\end{equation}
where $\ev{\bullet}_f$ denotes the fermion path integral. The mesonic two-point function is estimated by averaging Eq.~\eqref{eq:precorr} on gauge field configurations. At large distances, it decays exponentially with the energy of the lightest state with the specified quantum numbers
\begin{equation}
  C_\Gamma(x_0,y_0,\vb{p}) = \ev{ W_\Gamma(x_0,y_0,\vb{p}) } \propto \exp{-E_\Gamma(\vb{p})\abs{x_0-y_0}} .
\end{equation}
At the same time, the variance decays exponentially with the lightest state with at least four quark lines, i.e.\ a pair of zero-momentum pions, irrespective of $\Gamma$ and $\vb{p}$,
\begin{equation}
  \sigma^2\left[C_\Gamma\right](x_0,y_0,\vb{p}) = \ev{ \abs{W_\Gamma(x_0,y_0,\vb{p})}^2 }-\ev{ W_\Gamma(x_0,y_0,\vb{p}) }^2 \propto \exp{-M_{\pi\pi}\abs{x_0-y_0}} .
\end{equation}
Therefore, neglecting finite volume effects on the $\pi\pi$ state energy, any connected mesonic two-point function with the exclusion of pseudoscalar ones, e.g.\ $\Gamma=\gamma_5$, $\gamma_0\gamma_5$, has an expectation value that is decaying faster than its standard deviation. Assuming sampling over $n$ independent configurations with a standard MC, the \SN\ is
\begin{equation}\label{eq:sn_problem}
  \frac{C_\Gamma(x_0,y_0,\vb{p})}{\sigma\left[C_\Gamma\right](x_0,y_0,\vb{p})/n^{1/2}} \propto n^{1/2} \exp{-(E_\Gamma(\vb{p})-M_\pi) \abs{x_0-y_0}} .
\end{equation}

\section{The multilevel Monte Carlo algorithm}\label{sec:multilevel}

The multilevel algorithm addresses the \SN\ problem modifying the MC sampling and boosting the standard $n^{-1/2}$ scaling of the statistical error in Eq.~\eqref{eq:sn_problem}. This is achieved by splitting the MC estimator in two (or more) levels of sampling. At the lowest level-$0$, the observable is averaged on $n_0$ configurations that span the whole lattice gauge field. At level-$1$ (and higher), $n_1$ gauge field configurations are generated for each level-$0$ configuration. The level-$1$ updates are characterized by the independent update of different spacetime regions of the lattice, and they are used to independently average the observable in the distinct regions.

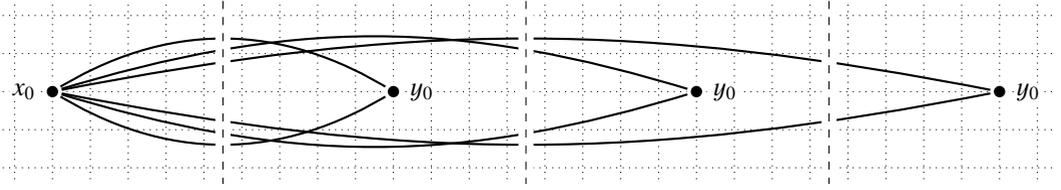
\begin{figure}[t]
  \centering
  \begin{tikzpicture}
    \begin{scope}
      \clip (-0.7,-1.2) rectangle (13.2,+1.2);
      \draw[step=0.5, dotted] (-1.5,-2.5) grid (14,+2.5);
      \draw[thick,bend left=32] (0,0) to ( 4.5,0) to cycle;
      \draw[thick,bend left=17] (0,0) to ( 8.5,0) to cycle;
      \draw[thick,bend left=11] (0,0) to (12.5,0) to cycle;
      \fill[white] (0,0) circle [radius=1.25mm];
      \fill        (0,0) circle [radius=0.75mm] node [left=1mm] {\contour{white}{$x_0$}};
      \foreach \x in {4.5,8.5,12.5}
      {
        \fill[white] (\x,0) circle [radius=1.25mm];
        \fill        (\x,0) circle [radius=0.75mm] node [right=1mm] {\contour{white}{$y_0$}};
      }
      \foreach \x in {0,4,8}
      {
        \draw[line width=2mm,white] (\x+2.25,-2.5) -- (\x+2.25,+2.5);
        \draw[dashed]      (\x+2.25,-2.5) -- (\x+2.25,+2.5);
      }
    \end{scope}
  \end{tikzpicture}
  \caption{Connected meson two-point functions in a two-level setup with multiple thick-time-slice regions.}
  \label{fig:cartoon}
\end{figure}

To this end, we divide the lattice in thick-time-slice regions, as pictured in Figure~\ref{fig:cartoon}. In general, we can write any global observable, such as $O=W_\Gamma(x_0,y_0,\vb{p})$ in Eq.~\eqref{eq:precorr}, as
\begin{equation}
  O = O_\text{fact} + O_\text{rest} , \qquad O_\text{fact} = O_0 O_1 \cdots O_m ,
\end{equation}
i.e.\ the product\footnote{
 Or, as in the case of $W_\Gamma(x_0,y_0,\vb{p})$, the contraction of the factors $O_i$ over tacit space and internal indices.
} of observables $O_i$, which are local to a thick time slice, plus a non-factorized correction $O_\text{rest}$. We introduced a way to build the factorized $W_\Gamma(x_0,y_0,\vb{p})$ from factorized quark propagators in Refs~\cite{Ce:2016idq,Ce:2016qto} for the two relevant regimes: when source and sink are both in the same region, and when they are in different regions. In both cases, $O_\text{rest}$ can be made small and it is easily dealt with using a standard algorithm. We focus here on connected contributions at large distances, so we are interested in the latter regime.\footnote{
  The disconnected contribution to a general fermionic two-point function shows an even stronger \SN\ problem. We successfully addressed the disconnected case using multilevel sampling in Refs~\cite{Ce:2016idq,Ce:2016ajy}.
} The multilevel estimator for $O_\text{fact}$ is
\begin{equation}
  \bar{C}^\text{mlv}_{\text{fact},\Gamma}(x_0,y_0,\vb{p}) = \frac{1}{n_0} \sum_{i=1}^{n_0} \left\{ \bar{O}_1 \bar{O}_2 \dots \bar{O}_m \right\} , \qquad \bar{O}_k = \frac{1}{n_1} \sum_{j=1}^{n_1} O_k[U_{i,j}] .
\end{equation}
Each factor $\bar{O}_i$ is estimated independently in the inner average and its statistical error scales with $n_1^{-1/2}$. If the statistical error of their product scales as $n_1^{-m/2}$, where $m$ is the number of regions that contribute to the observable, the standard error of the multilevel estimator scales with $n_0^{-1/2}n_1^{-m/2}$. Since the number of thick-time-slice regions is roughly proportional to the source-sink separation, $m\simeq\bar{M}\abs{x_0-y_0}$, the noise decreases with distance \emph{exponentially faster} with respect to the standard MC. We expect to have
\begin{equation}
  \frac{\bar{C}^\text{mlv}_{\text{fact},\Gamma}(x_0,y_0,\vb{p})}{\sigma\left[\bar{C}^\text{mlv}_{\text{fact},\Gamma}\right](x_0,y_0,\vb{p})/n_0^{1/2}} \propto n_0^{1/2} \exp{-\left[ E_\Gamma(\vb{p})-M_\pi-(\bar{M}\ln{n_1})/2 \right]\abs{x_0-y_0}} .
\end{equation}
Provided that is possible to boost the exponential rate of the noise to the level of the one of the signal, the \SN\ problem is completely solved. In general, the optimal value of $n_1$ is observable specific.

\section{The factorization of the lattice QCD action}\label{sec:mbddhmc}

In order to implement multilevel sampling in QCD with dynamical fermions, we need to be able to update the gauge field in different spacetime regions independently. To this purpose, in Ref.~\cite{Ce:2016ajy} we introduced a factorization of the fermion determinant and its dependency on gauge links in different regions.

\subsection{The domain decomposition}\label{sec:domdec}

As a first step, we set up a domain decomposition~\cite{Luscher:2005rx} in multiple active regions bordered by buffer regions with frozen gauge links, e.g.\ the decomposition of the lattice in even and odd thick time slices shown in Figure~\ref{fig:decomp}. Only the links in the $\Lambda_i$ with even $i$ are going to be active gauge links at level-$1$.

\begin{figure}[t]
  \centering
  \begin{tikzpicture}
    \begin{scope}
      \clip (-0.7,-0.6) rectangle (13.2,+1.1);
      \draw[step=0.5, dotted] (-1.5,-2.5) grid (14,+2.5);
      \foreach \x in {0,4,8}
      {
        \draw[dashed] (\x+1.75,-2.5) -- (\x+1.75,+2.5);
        \draw[dashed] (\x+2.75,-2.5) -- (\x+2.75,+2.5);
        \fill[pattern=north east lines,pattern color=gray] (\x+1.40,-2.5) rectangle (\x+1.70,+2.5);
        \fill[pattern=north east lines,pattern color=gray] (\x+2.80,-2.5) rectangle (\x+3.10,+2.5);
      }
    \end{scope}
    \foreach \x in {0,1,...,6}
      \draw (2*\x+0.25, 0.25) node {$\Lambda_\x$};
    \foreach \x in {2,4,6}
      \draw (2*\x-0.85,-0.25) node [font=\footnotesize] {\contour{white}{$\partial\Lambda_\x$}};
    \foreach \x in {0,2,4}
      \draw (2*\x+1.35,-0.25) node [font=\footnotesize] {\contour{white}{$\partial\Lambda_\x$}};
  \end{tikzpicture}
  \caption{Lattice decomposition in thick time slices. $\Lambda_i$ with even $i$ are update independently at level-$1$. $\Lambda_j$ with odd $j$ are frozen regions with thickness $\Delta$.}
  \label{fig:decomp}
\end{figure}

The determinant representing a single quark flavour is decomposed exactly in the product
\begin{equation}\label{eq:detQdd}
  \det Q = \frac{\det W_1}{\prod_\text{even $i$} \det{P_{\Lambda_i} Q_{\Omega^*_i}^{-1} P_{\Lambda_i}} \prod_\text{odd $j$} \det Q_{\Lambda_{j,j}}^{-1} } ,
\end{equation}
where $Q_{\Omega^*_i}$ is the hermitian Dirac operator $Q=\gamma_5(D_\text{W}+m)$ restricted to a three thick-time-slice region $\Omega^*_i=\Lambda_{i-1}\cup\Lambda_{i}\cup\Lambda_{i+1}$, $P_{\Lambda_i}$ projects on $\Lambda_i$, and we introduced the operator $W_z$, $z\in\complex$, defined as
\begin{equation}\label{eq:Wz_def}
  W_z = \begin{pmatrix}
    \dots & \dots \\
    \dots & z\Id & P_{\partial\Lambda_{i-2}} Q_{\Omega^*_{i-2}}^{-1} Q_{\Lambda_{i-1,i}} \\
    & P_{\partial\Lambda_{i}} Q_{\Omega^*_{i}}^{-1} Q_{\Lambda_{i-1,i-2}} & z\Id & P_{\partial\Lambda_{i}} Q_{\Omega^*_{i}}^{-1} Q_{\Lambda_{i+1,i+2}} \\
    & & P_{\partial\Lambda_{i+2}} Q_{\Omega^*_{i+2}}^{-1} Q_{\Lambda_{i+1,i}} & z\Id & \dots \\
    & & & \dots & \dots
  \end{pmatrix} ,
\end{equation}
and supported on the inner boundaries of active regions $\Lambda_i$, even $i$.

In Eq.~\eqref{eq:detQdd}, the determinant in each thick time slice is naturally factorized, but for a global residual contribution $\det W_1$. We argue that $\det W_1$ is parametrically close to one. To motivate this, we partition the active regions between doubly even ones $\Lambda_{\ell}$ and singly even ones $\Lambda_{\ell+2}$, where $\ell=4k$, $k\in\integers$. We define $\Id-w$ as the Schur complement of the singly even block of $W_1$. This leads to the identity~\cite{Jegerlehner:1995wb}
\begin{equation}
  \det W_1 = \det{\Id-w} , \qquad w = \left[ \sum_{\ell} P_{\partial\Lambda_{\ell}} Q_{\Omega^*_{\ell}}^{-1} Q_{\Lambda_{\ell\pm1,\ell\pm2}} \right] \left[ \sum_{\ell} P_{\partial\Lambda_{\ell+2}} Q_{\Omega^*_{\ell+2}}^{-1} Q_{\Lambda_{\ell+2\pm1,\ell+2\pm2}} \right] .
\end{equation}
The operator $w$ is supported on the internal boundaries of doubly even regions $\Lambda_{\ell}$ only, and it has a peculiar structure: the first factor propagates a quark line to the internal boundary of a singly even $\Lambda_{\ell+2}$ region, and the second factor propagates it back. This implies an exponential suppression with a rate proportional to $M_\pi/2$ times $2\Delta$, where $\Delta$ is the thickness of the buffer regions $\Lambda_j$ with odd $j$. A way to express this is that the eigenvalues $\delta_i$ of $w$ satisfy $\abs{\delta_i}\lesssim\order*{\bar{\delta}}$, with $\bar{\delta}=\e{-M_\pi \Delta}$.

Moreover, with algebra analogous to Eq.~(\href{https://journals.aps.org/prd/pdf/10.1103/PhysRevD.95.034503#temp:intralink-d13}{13}) in Ref.~\cite{Ce:2016ajy}, $w$ can be written as the product of two Hermitian operators. This implies that $w$ is similar to $w^\dagger$, thus the $\delta_i$ are either real or complex conjugate pairs. These constraints on the spectrum have been verified numerically in the simpler case of a three thick-time-slice setup, see Refs~\cite{Ce:2016ajy,Giusti:2017ksp} for the details.

\subsection{The multiboson representation}

The fact that the condition number of $\Id-w$ is $\simeq (1+\bar{\delta})/(1-\bar{\delta})=\order{1}$ suggests as a next step to apply to it the MB technique~\cite{Luscher:1993xx}, in its complex-valued variant~\cite{Borici:1995np}. This is obtained by approximating the operator inverse with a polynomial
\begin{equation}
  P_N(z) = \frac{1-R_{N+1}(z)}{z} = c_N \prod_{k=1}^N (z-z_k) , \qquad z_k = \{ z\neq 0 : R_{N+1}(z)=1 \} , \qquad u_k = 1-z_k .
\end{equation}
A simple and good choice is a truncated geometric series $P_N(z)=\sum_{p=1}^N (1-z)^p$, which has the nice property of being exponentially more accurate close to $z=1$, where the bulk of the UV eigenvalues of $\Id-w$ are concentrated. This polynomial choice has a constant accuracy on circumferences in the complex plane $\abs{1-z}=\text{const}$, but different choices might provide slightly better approximations.

Chosing $N$ even, we can write down the MB action for $\det{1-w}$
\begin{equation}\label{eq:mbfinal}
  \frac{\det{1-w}}{\det{\Id-R_{N+1}(\Id-w)}} \propto \prod_{k=1}^{N/2} \det{W_{\sqrt{u_k}}^\dagger W_{\sqrt{u_k}}}^{-1} \propto \prod_{k=1}^{N/2} \int \D*{\chi_k,\chi_k^\dagger} \e{-\abs{W_{\sqrt{u_k}}\chi_k}^2} ,
\end{equation}
where the MB fields $\chi_k$ are $N/2$ coloured Weyl spinor fields\footnote{
  It is worth noting that our MB representation applies naturally to a single quark flavour. Multiple flavours can be simulated with $N_\text{f}$ times more MB fields.
} supported on the inner boundaries of the active $\Lambda_i$. The derivation of Eq.~\eqref{eq:mbfinal} relies on the fact that $w$ and $w^\dagger$ are similar. Moreover, we used $\det\{z^2\Id-w\}\propto\det W_z$ to switch back to the operator $W_z$ defined in Eq.~\eqref{eq:Wz_def}. This last step is necessary to have a truly factorized MB action. Indeed, introducing $\chi_{i,k}=P_{\partial\Lambda_i}\chi_k$ as the MB fields on $\Lambda_i$, the MB action
\begin{equation}\label{eq:mbaction}
  \abs{W_z\chi_k}^2 = \sum_\text{even $i$} \abs{z\chi_{i,k} + P_{\partial\Lambda_{i}} Q_{\Omega^*_{i}}^{-1} \left[ Q_{\Lambda_{i+1,i+2}}\chi_{i+2,k}+Q_{\Lambda_{i-1,i-2}}\chi_{i-2,k} \right]}^2
\end{equation}
couples MB fields on $\Lambda_{i-2}$ to up to $\Lambda_{i+2}$. However, each term in Eq.~\eqref{eq:mbaction} depends only on the gauge links in a single active region $\Lambda_i$, and it does not depend on the gauge links in any other active region $\Lambda_{i'}$, $i'\neq i$. Because of this \emph{it is possible to update independently all the $\Lambda_i$ regions with even $i$}. The remaining factors in Eq.~\eqref{eq:detQdd} can be represented using ordinary algorithm based on pseudofermion fiels, restricting them to $\Omega^*_i$ or $\Lambda_i$ regions.

The resulting configurations are sampled with an action that differs from the Wilson action by a small reweighing factor $\mathcal{W}_N=\det{\Id-R_{N+1}(\Id-w)}$ for each dynamical quark flavour. However, given a reasonable $\Delta\approx\SIrange[range-phrase=\text{--}]{0.5}{1}{\fm}$, our tests show that $\lesssim 10$ MB fields for each light flavour are sufficient to obtain an a high-precision approximation, resulting in a negligible $\mathcal{W}_N$. We do not expect this figure to change significantly with lighter quarks.

\section{Numerical tests in the quenched theory}\label{sec:numeric}

Numerical tests are needed to asses the effectiveness of multilevel MC sampling in addressing the \SN\ problem of fermionic observables. To this purpose, we studied connected two-point functions such as the pion and baryon propagator~\cite{Ce:2016idq,Ce:2016qto}, as well as disconnected two-point functions~\cite{Ce:2016idq,Ce:2016qto,Ce:2016ajy}. In all these cases, multilevel sampling proved its value in solving or mitigating the \SN\ problem.

Here, we extend previous results considering two additional interesting observables:
\begin{itemize}
  \item the pseudoscalar correlator at non-zero momentum, which is a building block of transition matrix elements such as heavy meson decay form factors;
  \item the vector correlator, at zero momentum and averaged over $i=1,2,3$, which plays a prominent role in the computation of the hadron vacuum polarization (HVP) contribution to $g-2$ of the muon.
\end{itemize}

We tested the two-level sampling on these two-point functions in the quenched approximation. This has proven instrumental both to disentangle from the complications introduced by the factorization of the fermion determinant, and to limit the computational resources needed, while being still affected by the same \SN\ problem. The test was performed on $n_0=50$, $T\cross L^3=64\cross 24^3a^4$ level-$0$ configurations with open boundary conditions in the time direction, generated with the Wilson gauge action at $\beta=6$, which corresponds to $a\simeq\SI{0.093}{\fm}$ assuming $r_0=\SI{0.5}{\fm}$. The valence quark hopping parameter is set to $\kappa=0.1560$, which corresponds to $M_\pi\simeq\SI{455}{\MeV}$~\cite{Ce:2016idq}.

To keep things simple, we started with a two active thick-time-slice setup. For each level-$0$ configuration, we generated $n_1=30$ level-$1$ configurations updating independently the gauge field in $\Lambda_0=\{x:x_0\in(0,15a)\}$ and $\Lambda_2=\{x:x_0\in(24a,63a)\}$. Gauge links in $\Lambda_1=\{x:x_0\in(16a,23a)\}$ have been kept frozen.\footnote{%
  Although a frozen region as thin as a single time slice is sufficient to have independent quenched updates, we included a $\approx\SI{0.7}{\fm}$-thick frozen region to be consistent with dynamical fermions updates.
} For each active region, $n_1=30$ level-$1$ configurations have been used for measurements. A large number of MDUs have been skipped between measurements in order to reduce autocorrelation effects.

We computed the connected two-point function of the pseudoscalar density and of the local vector current using 12 spin- and colour-diluted random sources with $\mathrm{U}(1)$ noise, defined on the source time slice $x_0=8a$. Both sink and source were projected on momentum $\vb{p}^2=n(2\pi/L)^2$, with $n$ up to $3$. We show in these proceedings two specific correlators: $C_{\gamma_5}(x_0,y_0,\vb{p})$ with non-zero momentum $\vb{p}=2\cpi(1,1,0)/L$, and $C_{\gamma_i}(x_0,y_0,\vb{0})$ at zero momentum and averaged over $i=1,2,3$.

\subsection{Pseudoscalar correlator with momentum}\label{sec:pion110}

\begin{figure}[t]
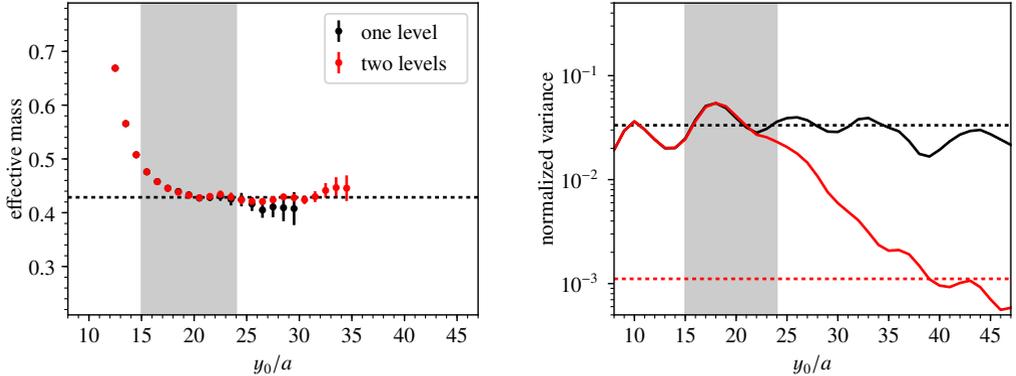

  \centering
  \scalebox{0.6667}{\input{./figs/proceedings_pion_110_effmass.pgf}}
  \scalebox{0.6667}{\input{./figs/proceedings_pion_110_relvar.pgf}}
  \caption{Plots of the effective mass (left) and normalized variance (right) of a pseudoscalar correlator with momentum $\vb{p}=2\cpi(1,1,0)/L$ on $n_0=50$, $n_1=30$ configurations, vs.\ the position $y_0$ of the sink operator. The source is at $x_0=8a$ and grey denotes the time slices not updated at level-$1$. The correlator is estimated with a standard estimator (black line) or with the two-level estimator (red line). The variance in the right plot is normalized to the variance with $n_1=1$.}
  \label{fig:pion110}
\end{figure}

The pseudoscalar correlator at zero momentum is a special case since already with a standard MC both the signal and the statistical error decay with the same exponential rate, so no significant \SN\ degradation with distance is expected. Our tests confirm that the multilevel algorithm results in no gain for the zero-momentum correlator. However, with non-zero momentum the \SN\ degrades with distance. For instance, with $\vb{p}=2\cpi(1,1,0)/L$ the \SN\ is expected to decrease with $E_{\gamma_5}(\vb{p})-M_\pi\approx 0.213/a$.

We show the results of the numerical computation in this case in Figure~\ref{fig:pion110}. In the right plot, the variance of the correlator is estimated on level-$0$ configurations after averaging on level-$1$ configurations, and it is normalized to the variance computed with $n_1=1$ in order to factor out the standard exponential suppression with distance. With $30$ level-$1$ configurations included but a standard single-level averaging procedure, the variance is $\approx 1/30$ as expected. On the other hand, with a two-level averaging procedure, the normalized variance depends on the sink $y_0$ coordinate. If both $x_0,y_0\in\Lambda_0$, the independent updates in $\Lambda_2$ do not affect the variance. If $y_0\in\Lambda_2$, the independent updates in the two regions result in a significant variance reduction. While at medium distances the variance reduction is saturated by less than $30$ level-$1$ updates, for $y_0\gtrsim 38a$ the normalized variance drops to $\approx 1/30^2$, in perfect agreement with expectations. We did not attempt to saturate the variance at longer distances pushing beyond $n_1=30$. In the left plot of Figure~\ref{fig:pion110} we show that the reduced variance translates into a smaller statistical error on the effective mass of the correlator at long source-sink separations. The signal stays constant for additional $\approx\SI{0.5}{\fm}$ before vanishing into the noise.

\subsection{Vector correlator}\label{sec:rho000}

\begin{figure}[t]
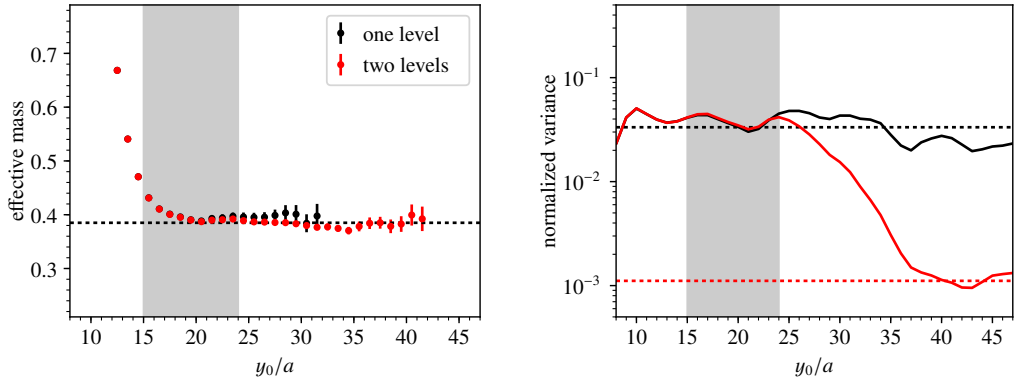

  \centering
  \scalebox{0.6667}{\input{./figs/proceedings_rho_000_effmass.pgf}}
  \scalebox{0.6667}{\input{./figs/proceedings_rho_000_relvar.pgf}}
  \caption{Plots of the effective mass (left) and normalized variance (right) of a vector correlator on $n_0=50$, $n_1=30$ configurations, vs.\ the position $y_0$ of the sink operator. The source is at $x_0=8a$ and grey denotes the time slices not updated at level-$1$. The correlator is estimated with a standard estimator (black line) or with the two-level estimator (red line). The variance in the right plot is normalized to the variance with $n_1=1$.}
  \label{fig:rho000}
\end{figure}

In the setup considered here, the vector correlator decays asymptotically with the mass of the $\rho$ meson state. Since the noise is reduced only according to $M_\pi$, the vector correlator at zero momentum suffers a exponential \SN\ problem with a rate $M_\rho-M_\pi\approx 0.170/a$.

We addressed this problem with the multilevel. As in the previous case, in the right plot of Figure~\ref{fig:rho000} we show the normalized variance of the correlator, with and without multilevel averaging. The same considerations of Section~\ref{sec:pion110} apply: the two-level averaging procedure results in a variance perfectly scaling with $n_1^2$ when source and sink are in different regions. As shown in the left plot of Figure~\ref{fig:rho000}, the two-level estimator allows the signal in an effective mass plot to be followed for additional $\approx\SI{0.9}{\fm}$ with respect to the standard estimator, opening new perspectives for the determination of the HVP contribution to $g-2$ of the muon.

\section{Conclusions}

We showed in Ref.~\cite{Ce:2016ajy} that it is possible to factorize the gauge field dependence of the quark determinant, introducing a domain decomposition in overlapping domains and representing the residual contribution with multiboson fields. Combined with the quark propagator factorization introduced in Ref.~\cite{Ce:2016idq}, the quark determinant factorization allows multilevel sampling of fermionic correlators, for the first time with dynamical fermions. In these proceedings, we focused on a setup with multiple independent thick time slices, that results in the action in Eq.~\eqref{eq:mbaction}.

The effectiveness of two-level sampling has been tested in Refs~\cite{Ce:2016idq,Ce:2016qto,Ce:2016ajy} for the baryon propagator and for the disconnected contribution to the flavour-singlet pseudoscalar propagator. In these proceedings, we tested in addition the two-point functions of pseudoscalar densities at non-zero momentum and of vector currents in a two-region setup. Results show an exponential gain in the \SN\ in both cases. This suggests new ways to approach computations such as the hadronic vacuum polarization contribution to the anomalous magnetic moment of the muon.

Variants on the combined domain decomposition and multiboson scheme might also address the limitations of standard HMC algorithms in master field simulations with dynamical fermions~\cite{Luscher:2017cjh}.

\section*{Acknowledgements}
Simulations were performed on Galileo and Marconi at CINECA (CINECA-INFN and CINECA-Bicocca agreements). We thank these institutions for the computer resources granted.

%
\bibliography{biblio}

\end{document}